\documentclass[a4paper,11pt]{scrartcl}

\usepackage{latexsym}
\usepackage{times}
\usepackage[latin1]{inputenc}
\usepackage{url}

\newif\ifpdf
\ifx\pdfoutput\undefined
\pdffalse
\else
\pdfoutput=1
\pdftrue
\fi

\ifpdf
\fi

\author{Maximillian Dornseif\thanks{Laboratory for Dependable
    Distributed Systems, RWTH Aachen University -- dornseif@informatik.rwth-aachen.de} \qquad
Sascha A. May\thanks{sascha@may.de} 
}

\title{Modelling the costs and benefits of Honeynets}
\date{\today}

\begin{document}

\maketitle

\begin{abstract}

For many IT-security measures exact costs and benefits are not known. 
This makes it difficult to allocate resources optimally to different security measures. We present a model for costs and benefits of so called Honeynets. This can foster informed reasoning about the deployment of Honeynet technology.

\end{abstract}

\section{Introduction}

Honeynets are collections networked of computer systems which are
intended to be attacked and broken into in an observed fashion,
keeping track of any (mis-)use. 
Similar to other IT-security technologies there is a lot of
gospel on the benefits of Honeynets, while there is little analysis on the
exact gain offered by them and the associated cost. We are presenting a
model helpful in understanding the economic aspects of Honeynet
deployment.

In section 2 we present an overview of the technical aspects of Honeynets.
Sections 3 and 4 collect benefits and respective costs of Honeynets.
Following that section 5 contains our model of Honeynet deployment
followed by section 6 where we summarize our findings.

\section{Honeynet Technology}

Honeynets are a term used to describe one or more computers destined as
being penetrated and supportive technology designed to capture activity on
the Honeynet and to decrease the risk imposed by the Honeynet to other
systems \cite{project01:_know_your_enemy}.

The usual setup consists of the hosts to be attacked connected via a
transparent firewall to the Internet. This firewall termed ``Honeywall''
is responsible for logging all network traffic entering and leaving the
Honeynet. It also tries to suppress grave attacks from compromised hosts
in the Honeynet by rate limiting outgoing traffic radically and by
rewriting outgoing traffic on the fly to stop known exploits initiated
from the Honeynet.

The hosts to be attacked are instrumented in a way aimed at allowing the
operators of the Honeynet to capture encrypted network traffic on the host
where it exists in decrypted form while users of this hosts. This is either archived by using
trojaned binaries like SSH which log data processed by them in
disguise or by modifying the kernel to log certain
system calls resulting in a crude keystroke logging facility. The captured
data is covertly transferred to the Honeywall for logging.

\section{Benefit of Honeynets}

Deployment of Honeynets results in information gathered and possibly an
increased security for the operator of the Honeynet.

\subsection{Possible Information Gain on Attacks by Honeynets}

Honeynets can gain information on the attacks against them. We assume
that a Honeynet  can basically gather two different qualities of
Information: After starting his attack at $t_a$ the attacker is unaware of
the fact that he is attacking a Honeynet the data gathered shows the
attacker's typical
actions against the class of system the Honeynet is emulating. At a
certain point in time the attacker realizes that he is confronted with
a Honeynet. At this point labeled $t_d$ the attacker's motivation
shifts which should also result in a change of behavior. $t_d$ can be even
before $t_a$ if an attacker is able to gather
information about the Honeynet out of band and attacks with the
knowledge that he is attacking a Honeynet. $t_d$ also can be in infinitive
future if the attacker isn't willing or able to find out that he is
attacking an
Honeynet.

It is safe to assume that after $t_d$ the attacker will be more
reluctant to act in a way which will allow the observer to gather
further information. The attacker usually will completely stop the attack
and
vanish. But we also know of one instance where attackers using the
Honeynet as an IRC proxy just ignored the fact that they where
observed.

While attacking the attacker will try to escalate his privileges. He will
increase his privileges in zero or more steps. The higher he was able to
escalate his privileges the more likely he is to find out the true nature
of the host he is attacking which results in  $t_d$ moving into future.

It is therefore safe to assume that sophisticated attackers $t_d$ is
relatively early. \cite{dornseif04:_noseb_attac_honey} A sophisticated attacker will be able to escalate
his privileges relatively fast increasing his chances of detection.
For attackers with full local privileges detecting a Honeynet is
trivial.

Honeynets can not collect informations on all kind of attackers equally.
Honeynets are be able to gather representative data on attackers
which choose their targets more or less randomly like autonomous malware
and very unsophisticated attackers do. Gathering data more on focused
attackers can be only done for attackers actively choosing to attack the operators systems.

An attacker not penetrating systems in a random fashion must be tricked
into attacking a Honeynet by making it look like a worthwhile
target. It can be assumed that the more sophisticated the attacker is
the less likely he will fouled by such deceptions.

So while Honeynets might be able to gather relatively much Information
about unsophisticated attackers or autonomous malware like worms, with
the same investment much less Information can be gathered about
sophisticated attackers.


\subsection{Possible increased security by using Honeynets as an decoy}

It is claimed that Honeynets can increase the search space for finding
valuable systems in a network and thus increasing security by
luring attackers into spending effort attacking the Honeynets instead
of the real thing. This claim has to be evaluated against different
adversary scenarios.

Attackers attacking random hosts in your network have a bigger search
space. But only extremely unsophisticated attackers like autonomous
malware can be assumed to attack completely random hosts. Also these
attackers can only be significantly slowed down when a significant
percentage of a network are Honeynets which is unlikely.

More sophisticated attackers will choose their target based on their
objectives and on a systems perceived value to complete this
objectives. Simply by their existence Honeynets will slow down the
attackers
target selection process. To foul the attacker in attacking the
Honeynet the Honeynet has to look more attractive than the target the
attacker is aiming for or the ``real'' system has to be hidden in a
way that the attacker will not be able to detect it.

\subsection{Possible increased security by using aggressive Honeynets for redirection}

There are also attempts to deploy honeypots as part of active
network security. It is tried to reroute attackers from a production
server to a Honeynet  distracting the attacker and allowing further
gathering of data \cite{whitsitt:_bait_switc_honey}.

The detection of the attack triggering the rerouting is a non trivial
problem. Also the Honeynet must mirror very closely the production host to
make rerouting seamless and less detectable.
Due to this unsolved problems we will exclude aggressive Honeynets from
further investigation.

\section{Costs of Honeynets}

Honeynets come with costs. Costs can be separated into costs for deploying
and operating the Honeynet and costs due to increased risk imposed by the
Honeynet to the operators network.

\subsection{Cost of deploying}

The initial costs of deploying a Honeynet at $t_0$ consists of the
hardware,
including computers, network devices and wiring, housing, personal cost
for setup and cost of fitting the Honeynet into the policy framework of
the organization. Since Honeynets are a relatively little understood and
very new technology, personal costs for setup are likely to be
exceptionally high. Also since Honeynets are something relatively new,
explaining them to all stakeholders in the organization, evaluation of
policy implications and weighing risks against benefits is likely to
consume considerable resources.

\subsection{Cost of operation}

Operational costs consist of maintenance costs consist of fixed costs for
housing, power, basic monitoring and software maintenance. Considerable
maintenance effort has to go into maintaining the Honeywall to ensure the
security of itself and capability to minimize the risk of being used as a
stepping stone to attack further systems.

Between $t_a$ and $t_d$ variable costs arise for IP-traffic, log space,
forensics and active avoidance of being used as a stepping stone to attack
further systems. At $t_d$ variable costs for damage repair arise.

Since in the Internet there is climate of permanent attack, $t_0$ and
$t_a$ are likely to coincide. This results in permanent incurrence of the
variable costs.

\subsection{Cost of increased risk to own network}

By deploying a Honeynet the complexity of a network is increased
\cite{spitzner03:_honey_farms}. Also less-than state of the art protected
systems are added to the network. This decreases network security and
legal risks to the organization running a Honeynet.

\subsubsection{Risk to your Security}

The obvious effect of adding complexity and not state of the art protected
systems is decreased security resulting in increased risk to the network
attached to the Honeynet. Possible scenarios include the Honeynet
attracting additional attackers, unexpected interaction of the Honeynet
with other network components, use of the Honeynet as an attack platform
against others despite countermeasures implemented in the Honeywall or the
own network or use of the Honeynet for gathering intelligence about the
attacked organization and it's methods.

\subsubsection{Legal Liability Risk}

The main liability risk of Honeynets consist of the risk that an attacker
uses the Honeynet to attack systems of a third party and that this party
seeks damages against the operator of the Honeynet. Legal issues raised by
Honeynets have seen up to now no in-depth analysis by experts.  While the
whole ``downstream liability'' issue, that is liability of organizations
whose IT-systems are penetrated and used to attack others, has seen no
consensus or even satisfying discussion by the legal community, this
discussion is often use to argue that there is no liability to operators
of Honeynets \cite{spitzner03:_honey_legal}.
Such argument misses the important fact that the ``downstream liability''
is about possible negligence in securing your systems against misuse by
third parties, while with Honeynets you are actually willingly and
knowingly facilitate your network to be misused by making it less secure.

This carries a risk of legal liability against the operators of a Honeynet
 not only in relation to possible victims of an attack but also
possibly liability of management in relation to the shareholders of the
organization operating the Honeynet.  Besides civil liability there might
also be criminal liability depending very much on the jurisdiction
applied.

\section{Modelling}

Based on this observations we can build a microeconomic model of
Honeynets:

In trying to build a model of Honeynet operation we assume that during the
attack between $t_a$ and $t_d$ the attacker does a move every unit of
time. He manages zero or more times to escalate his privileges. Regardless
of that every unit of time the Honeynet operator earns the same value of
information on the attacker.

\subsection{Honeynet Operator}

We model the Honeynet operator according to the following rules:

\begin{enumerate}

\item The operator of the Honeynet is not interested in observing attacks
per se, but only on attacks specifically aimed at his systems, since
information on less focused attackers can be bought on the marketplace.
The attacker which is of interest to the operator will be called qualified
attacker and probably has the profile of a professional spy. We assume
that information on qualified attackers is not available in the
marketplace.

\item We assume that only a an extremely small percentage of attacks are
committed by qualified attackers.

\item We assume an attacker stops the attack after he discovers the nature
of the Honeynet. So after $t_d$ there will be no additional attacker
activity.

\item At $t_0$ the Honeynet is deployed and generates considerable fixed
startup costs.

\item Every unit of time the Honeynet generates constant costs for housing,
energy routine maintenance and updates.

\item Between $t_a$ and $t_d$ costs for additional resources like
bandwidth and logfile storage, monitoring, increased risk, restoring of
the Honeypot and forensic analysis arise. While some of this costs are per
unit of time during an attack, others occur only at the end of an attack.

\item By investing in the Honeynet the Operator can make the true nature
of the System harder to detect thus moving $t_d$ to the right.

\item For every unit of time the attack by an qualified attacker persists,
the operator gains information of a certain constant value. The
information gained while attacks by unqualified attackers is worthless.

\item On the Internet there is a climate of permanent aggression. This
means that $t_0 = t_a$ and that the constant stream of attacks can be
superimposed to build a constant attack pressure.

\item The Operator is not interested in prosecuting attackers
  \cite{honey_projec}. Experience shows that companies are extremely
  reluctant to prosecute computer crime. Also the circumstances under
  which evidence is gathered in Honeynets suggests that using such
  evidence in court would be difficult.

\end{enumerate}

We feel that this rules can realistically model most potential operators
of a Honeynet. A prominent exception are institutions whose mission
includes information security or criminology research. To these
organizations qualitative and quantitative data not only on qualitative
attackers but on any attack might be of value.

\subsection{Attacker}

We model the attacker according to the following rules:

\begin{enumerate}

\item The attacker has fixed costs like office space and Internet access.

\item The attacker has fixed costs per unit of time during the attack
starting at $t_a$.

\item The possible duration of the attack is unlimited and can converge to
infinity.

\item The attacker learns at $t_d$ that the attacked host is a Honeynet.

\item The Honeynet can not be perverted to remove collected information on
the attacker. A qualified attacker will not use it as an attack
proxy to third party systems.

\item Since the attacker has minimal interest on fake data, to the
attacker the Honeynet suddenly turns out as worthless. The Attack is being
ended.

\end{enumerate}

These rules we use for modeling the attacker are a profound restriction on
the types of behavior in the real world by assuming that the Honeynet can
not be perverted. Real world experience shows that a seasoned attacker
might be able to pervert all components of a Honeynet and use them to his
will.

Also the costs to the attacker are hard to estimate. One could speculate
that an attacker will use the cheapest forms of attack first and resort to
more costly attacks gradually when failing to reach it goals with the
cheaper attacks. Than again the attacker might set priority to avoiding
detection and use his best, most expensive tools attacks at first to do
so. This could be modeled by including the specific risk of detection in
the cost of an attack. We argue that most attacks are in a the same
magnitude and that the attacker usually does not base its decision on cost
of a specific attack but on his actual level of intelligence and
perception which attack might be most successful. Therefore we assume a
fixed costs per unit of time during the attack.

\section{Composition}

Building on those assumptions we can construct two curves describing the cost and the utility associated with a Honeynet:
  
\begin{enumerate}

\item The cost of a Honeynet is expressed by $c(t) = S + Mt$ where $S$ are the startup cost for deploying the Honeynet and $M$ are the maintenance costs per unit of time. Maintenance costs are bound to be greater than zero.

\item The utility of a Honeynet is expressed by $u(t) =
  Pt\frac{M}{I}$ whereas $P$ is the value of information
  gained by a single attack and $I$ is the factor by which higher
  investments in the maintenance influence the likelihood of being
  attacked.


\end{enumerate}

The use of a Honeynet is profitable if the curve $u(t)$ supersedes $c(t)$. If we vary $M$ while keeping $S$, $P$ and $I$ constant we can find the optimum investment in maintenance for organizations using Honeynets.

\section{Conclusion}

While the start-up cost $S$ seems relatively easy to determine, $P$ and $I$ are problematic. $P$ refers to the value of the Information gained by observing a qualified attacker. Since this information is not available on the marketplace it's technical value is extremely high. But we doubt that the actual utility of this information to the operator is that high in all cases. Sometimes actions by qualified attackers might only qualify as boring.

$I$ expresses the relation between maintenance and frequency of qualified attacks. 
We are not aware of means to determine real-world values for $I$ but we suspect that the general frequency of attacks is very low and can be only slightly increased by investing more.

\bibliographystyle{plain}
\bibliography{honeyeco}
    
%




\end{document}